\def\etal{{\it et al.\/}\ }

\def\eg{{\it e.g.\/}\rm,\ }
\def\lsim{~\rlap{$<$}{\lower 1.0ex\hbox{$\sim$}}}
\def\gsim{~\rlap{$>$}{\lower 1.0ex\hbox{$\sim$}}}
\def\void#1{{}}

\documentclass[runningheads]{cl2emult}

\usepackage{makeidx}  
\usepackage{graphicx} 
\usepackage{subeqnar} 
\usepackage{multicol} 
\usepackage{cropmark} 
\usepackage{eso}      
\makeindex            

\def\eg{{\it e.g.}}

\begin{document}
\title*{Public Imaging Surveys: Survey Systems and Scientific Opportunities}
%
%
%
%
\titlerunning{Public Imaging Surveys: Scientific Opportunities }
%
\author{Luiz da Costa}

\authorrunning{Luiz da Costa}
%
%
\institute{European Southern Observatory, Karl-Schwarzschild-Str{\ss}e 2,
 Garching bei M\"unchen, D-85728,  Germany}

\maketitle              

\begin{abstract}

The start of operation of several large-aperture telescopes has
motivated several groups around the world to conduct deep large imaging
surveys, complementing other wide-area but shallower surveys. A special
class of imaging surveys are the public ones such as those being
conducted by the ESO Imaging Survey (EIS) project established with the
primary objective of providing to the ESO community large datasets from
which samples can be drawn for VLT programs. To date several surveys
have been carried out providing multi-passband optical data over
relatively large areas to moderate depth, deep optical/infrared surveys
of smaller fields as well as observations of a large number of selected
stellar fields.  These surveys have produced a large amount of
well-defined datasets with known limits from which a variety of data
products have been extracted and distributed. The goal of the present
contribution is to give an overview of the EIS project, to present the
end-to-end survey system being developed by this program and to highlight
the effort being made to standardize procedures, to build software tools
to assess the quality of the derived products and to explore
the available datasets in a systematic and consistent way to search
for astronomical objects of potential scientific interest.

\end{abstract}

\section{Introduction}

The start of regular operations of several 8-m class telescopes has
created a new demand for large imaging surveys.  However, in contrast
to other ongoing all-sky surveys, there is a pressing need for
customized surveys designed in support of programs to be conducted
using large-aperture telescopes. These surveys should be capable of
providing data sets from which samples suitable to the instrumentation
available on these telescopes can be drawn, in particular reaching
magnitudes comparable to their spectroscopic limit.  Currently, most
of the major observatories are involved in large optical/infrared
imaging surveys using either dedicated 2-m class telescopes or
generous amounts of time on 4-m telescopes. These surveys are being
carried out either by individual groups or in the form of public
surveys, thereby serving a broader community and increasing the
scientific yield of the accumulated data.

The need for imaging data is especially acute for the ESO community at
large, considering the four telescopes of VLT and the large array of
spectrographs, some lacking imaging capabilities. Foreseeing this
need, ESO established in 1997 the ESO Imaging Survey (EIS) project
with the primary purpose of conducting public imaging surveys in
support of VLT science.  Over the past three years EIS has carried out
several surveys providing the ESO community with multi-passband
optical data over relatively large areas to moderate depth, deep
optical/infrared surveys of smaller fields as well as observations of
a large number of selected stellar fields.  Some of these programs
cover sky regions being observed from the ground and space in other
wavelengths (\eg~Chandra and HDF-S fields). Combined, they provide
valuable multi-wavelength datasets from which statistical samples can
be extracted and searches for rare population of objects can be
conducted. In fact, the imaging surveys driven by 8-m class telescopes
sample a region of parameter space significantly different from that
accessible by shallower, all-sky optical/infrared surveys, thus
offering new scientific opportunities for a broad range of areas.

In this paper the requirements for establishing a framework for
conducting long term ground-based, optical/infrared public surveys and
for designing a robust end-to-end survey pipeline are reviewed.  The
broad range of issues that impact the architecture of a survey system
are identified in section~2. In section~3, the EIS project is briefly
reviewed and used as a concrete example of an ongoing public survey.
This experience has served to set more specific requirements on the EIS
survey pipeline software presented in section~4. This system was
specifically designed to meet the needs of long-term, multi-instrument
public imaging surveys carried out by a small team. In this section some
of the main features of the new EIS pipeline architecture are described
as well as some other important components of the system and its current
status. A more detailed discussion of some of the issues involved in the
preparation of derived products is presented in section~5, followed by a
brief summary in section~6.

\section{Public Imaging Surveys}

The need for imaging surveys is of course not new - what has
dramatically changed in the past few years is the rapid increase in the
rate digital images are being collected, and consequently the volume of
data to be processed, and the time pressure to convert these data into
input samples for follow-up observations. The challenge is now to build
efficient systems to store, process, access, visualize, analyze and
scientifically exploit the resulting large data sets.  The basic
reduction steps are well-known and include: removal of instrument
signature; astrometric and photometric calibration; source detection and
measurement of flux and structural parameters; and object classification
(star/galaxy). Less standard steps include the combination of
information obtained using different passbands and/or telescopes,
stacking and mosaicing of images, preparation of color catalogs and use
of multi-wavelength information for object classification.

Depending on the duration of the specific program it is desirable to
consolidate these various steps into a pipeline enabling the efficient
and un-supervised reduction of the data. The need for pipelined
operations is particularly strong for long-term programs for which a
greater degree of autonomy and a more complex data flow model are
essential to ensure efficiency, uniformity and consistency in the
production of deliverables. The pipeline requirements may vary
considerably depending on the {\it nature} (public/private) and the
{\it size} of the survey program. It is important to emphasize that not
all {\it pipelines} are or need to be alike. In fact, the design of a
survey pipeline depends on a variety of factors including: the specific
goals of the survey and its strategy (number of passbands, area, depth,
type of fields, dithering pattern, mosaic geometry); expected input data
rate and total data volume; the number of telescope/detector
configurations involved; the variety of derived products; and how
frequently intermediate and final products should be made available. One
must also consider the desired level of uniformity of the data in terms
of  photometric accuracy and seeing, which may vary depending on the
specific scientific application.  All of these points translate into
requirements for the design of the survey software which must be able to
store and administrate all this information and process the data
according to their specifications. Of course, the larger the set of
requirements is the  more  complex the  pipeline needs to be, something
not fully appreciated. It must be emphasized that the development of a
survey system only pays dividends in the long-run.

To be successful a public survey depends not only on the design of the
associated survey pipeline but also on the context in which it is
inserted, as it must rely on the infrastructure available and the level
of support granted to the program. A public survey must rely on a
variety of interfaces and a balance must be reached between the
specific responsibilities of the survey team and the institute at
large. Some important points to consider are: whether the telescope
used is a survey-dedicated or a general-purpose one; the nature of the
telescope/survey team interface; the availability of tools to prepare
and monitor observations; the availability of calibration data; how
efficiently raw data can be accessed by the survey team as well as
their overall uniformity.

Since the main goal of a public survey is to have data reach the
interested users, one must also make sure that tools and resources are
available to distribute final products and to provide some level of
on-line service. The latter is particularly important since the
present technology precludes the widespread distribution of pixel
maps. In the short-term the next best alternative is to provide as
many derived products as possible in the form of source catalogs,
while allowing users to at least retrieve image sections to assess by
themselves the quality of the delivered products for their scientific
goals. This also provides the opportunity for independent verification
of the data and a possible way to obtain an important feed-back from
the users, leading to further improvements in the preparation of final
products. Another mid-term alternative is to distribute the data to
national/regional data centers which can then further distribute them
to individual institutes and university research groups.

An additional complication for ground-based optical/infrared surveys is
that they are, in contrast to surveys carried out from space, sensitive
to weather and atmospheric conditions. This implies that the survey data
have always to be supplemented with information regarding seeing and
photometric quality. For this purpose, tools must be built to
automatically monitor the progress of the survey not only in terms of
its completeness but also of the quality of the data. More importantly,
this has to be done quickly to enable the preparation of new
observations, including repeat observations, and  to prepare periodic
updates to inform the general public of the status of the survey. In
addition, when a dataset is released this information should be properly
summarized in supporting publications to fully characterize the block of
data being distributed. This is particularly problematic when releasing
incomplete data sets, which may be the case when a pre-defined schedule
of data releases is adopted.

Careful consideration of the several issues mentioned above are
essential to establish the requirements for a survey pipeline software
and the general framework within which a long-term, sustainable public
survey can be carried out.

\section{The EIS Project}

The EIS project was established by ESO with the short-term goal of
providing the ESO community different samples for the scientific
verification of the first unit telescope of the VLT and its first year
of operation. It was an attempt to coordinate and maximize the return
of foreseen preparatory imaging programs at the NTT by consolidating
them into a single survey and making the data immediately available to
the ESO community. Besides this short-term goal, the EIS project was
also viewed as an important first step to: establish a framework for
conducting public surveys at ESO; drive the development of software to
be consolidated into a survey system; create mechanisms to tap on the
expertise spread throughout the ESO community; disseminate know-how by
training students and post-docs and by making available survey
software and tools; and create an environment for testing and
bench-marking survey system software using real data.

From the organizational point of view, the first step was to appoint a
Working Group assembled from scientists, representing a broad range of
scientific interests, drawn from the community and in charge of:
designing and supervising the progress of the surveys; preparing
proposals and progress reports to the time allocation committee;
proposing to the ESO management policies regarding data distribution
and timetables; acting as representatives of the scientific interests
of the ESO community at large. In parallel, a special Visitor Program
was created to fund short- and long-term visits to ESO of known
experts in the field as well as students and post-docs to form the
Survey Team. The responsibilities of the team include: 1) preparation
and execution of observations; 2) data reduction; 3) production and
verification of calibrated images, object catalogs and targets; 4)
maintenance of supporting database; 5) maintenance of web pages; 6)
periodic reports describing the progress of the observations and
papers accompanying each data release describing the data, the
methodology adopted in the reduction and in the preparation of derived
products. In the particular case of EIS, all these tasks have been
carried out in parallel.

\begin{figure*}
\centering
\includegraphics[width=0.75\textwidth]{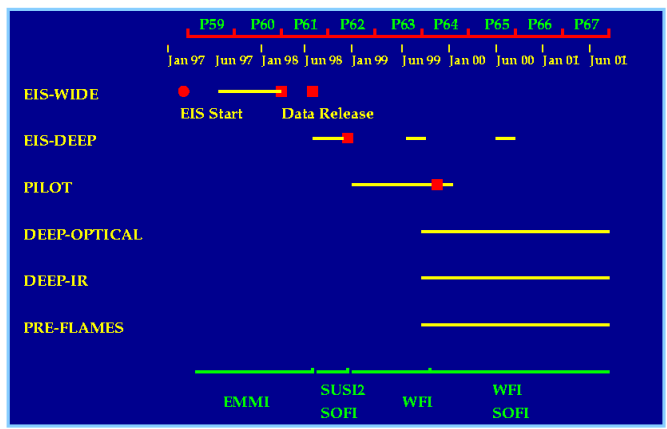}
\caption[]{Timeline of the EIS project indicating for each survey,
the period of observations and the dates data were publicly released.
Also shown, at the bottom of the figure, the different instruments
used in the different surveys. }
\label{fig:timeline}
\end{figure*}

\begin{figure*} \centering
\includegraphics[width=0.75\textwidth]{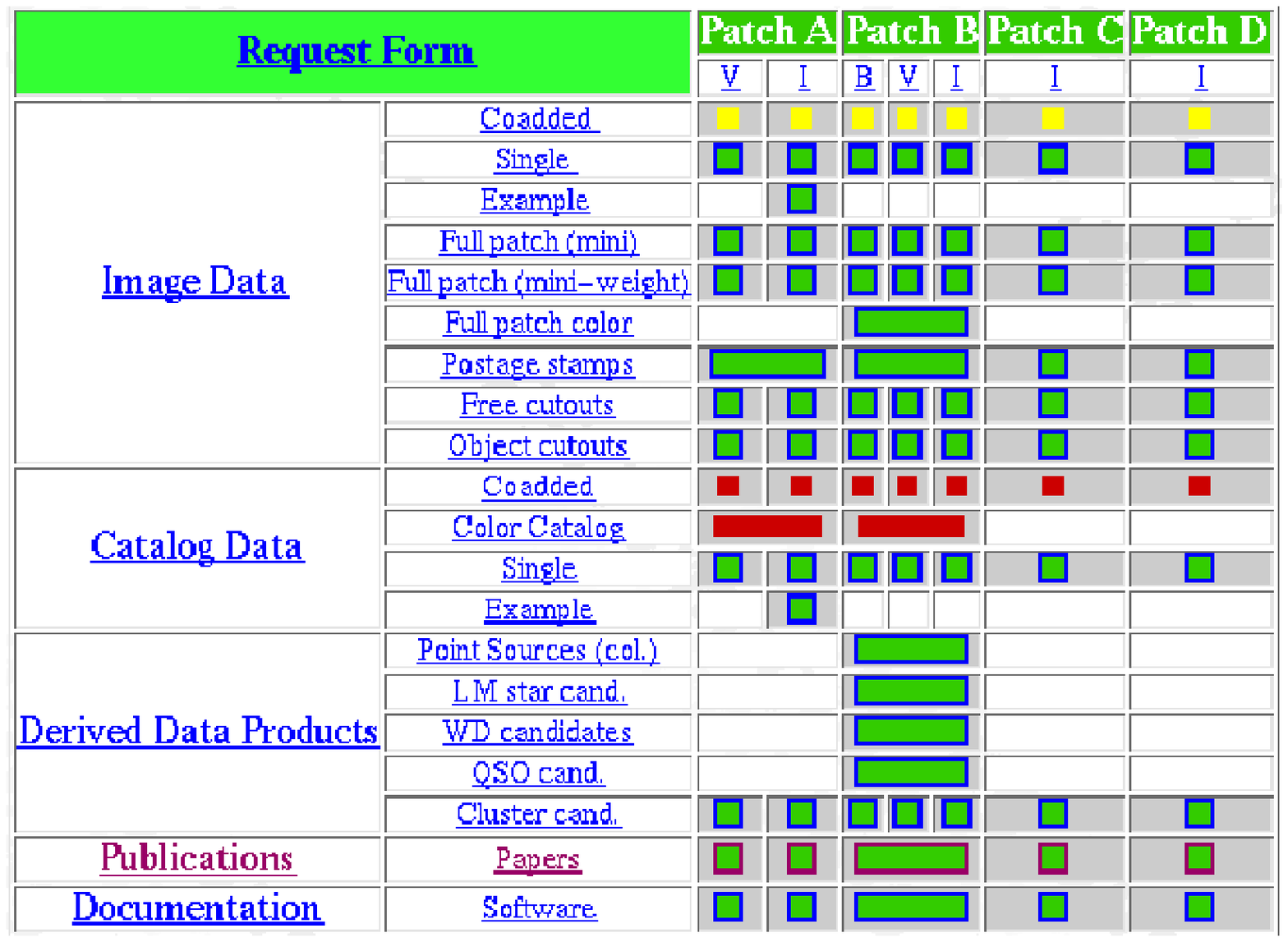}
\caption[]{EIS data request form available on the web.} \label{fig:release} \end{figure*}

As pointed out above, EIS started as a short-term experimental project
and in its early phase a pragmatic approach was adopted to develop a
basic pipeline. The main effort was to assemble and interface existing
software (\eg~IRAF, Drizzle, LDAC, SExtractor) to enable raw images to
be processed and astrometrically calibrated, mosaics created and
object catalogs produced under a minimum of supervision for the
specific instrument and survey strategy adopted in the first
survey. Due to time limitations much of the administrative tasks,
photometric calibration, data verification and archiving remained
external to the pipeline proper and required human intervention. This
first implementation of the pipeline was effective for the input data
rate and the total volume of survey data expected at the
time. However, the pipeline could not match the requirements imposed
by the rapidly evolving goals of the imaging survey, as can be seen in
Figure~\ref{fig:timeline} where the timeline of project is shown. This
figure shows: at the top the dates and ESO's observing periods; on the
left side the various surveys conducted by EIS; and at the bottom the
different detectors (telescopes) used. Also shown in the figure are
the periods of observations of each survey and the dates of previous
data releases. For reference, the science verification observations of
VLT UT1 were carried out in December 1998, and included targets
provided by EIS. The public nature of EIS requires a wide range of
data products to be made publicly available periodically in the form
of incremental and complete data releases \cite{renzini}.  This is
illustrated in Figure~\ref{fig:release} which shows the web page from
which EIS data products can be requested.  This particular example
corresponds to the complete data release of the EIS-WIDE survey. Given
the large variety of products that are made available it is critical
to have in place standard and automatic procedures to prepare these
releases, to ensure the quality and consistency of the products.
These requirements have a strong bearing on the scope of the EIS
survey system and on its design.

From Figure~\ref{fig:timeline} one can see that over the past three
years six different surveys have been conducted (more information on
the EIS project can be found at the URL www.eso.org/science/eis):

\begin{enumerate}

\item  EIS-Wide: a moderately deep,  wide-angle survey conducted at
the NTT. The survey covers 17 square degree in $I$-band as well as
smaller areas in $B$ and $V$.

\item EIS-DEEP: a multi-passband optical/infrared deep survey of the
HDF-S and Chandra (AXAF) fields.

\item Pilot Survey: a $BV$ wide-angle survey covering over 14 square
degrees, conducted with the wide-field imager (WFI) mounted on MPG/ESO
telescope, to complement the $I$-band survey.

\item  Deep Optical Survey: a deep $UBVRI$ optical survey
intended to cover three, one square degree strips (four WFI pointings)
over a range of right ascensions. Even though it can be considered as
an extension of EIS-DEEP it is based on observations being carried out
with WFI.

\item Deep Infrared Survey: designed to cover two areas of 450~square
arcmin in two patches also observed in the optical.

\item  Pre-flames: a multi-passband, relatively shallow survey,
of selected stellar fields to be used for observations with the
FLAMES/Giraffe combination.

\end{enumerate}

In summary, the various EIS surveys were designed to provide datasets
for a wide range of science: multi-passband data over a relatively
wide-area to moderate depth to search for rare objects (\eg~
high-redshift clusters, quasars), deep surveys for the selection of
statistical samples of galaxies for high-redshift studies (\eg~LBG)
and imaging data with accurate astrometry ($\lsim~0.2$ arcsec)
for multi-object spectroscopy using the high-resolution Giraffe
spectrograph, lacking imaging capabilities.  The selected pointings
have also been obvious choices. The wide-angle surveys consisted of
four distinct patches covering a wide range of right ascensions, to
provide targets almost year-round, in a declination range suitable for
VLT observations. Two of the selected regions are located at
high-galactic latitudes: one near the South Galactic Pole, of interest
for studies of quasars and absorption systems, and the other
overlapping with ESO Slice Project redshift survey and the associated
deep ATESP radio survey~\cite{prandoni}. The other two regions
surveyed by EIS-WIDE were selected at lower galactic latitudes to
provide data of potential interest for galactic studies. Three of the
selected regions have also been covered by radio surveys such as
ATESP, NVSS and WISH. The deep surveys have included regions where
considerable work is under way in other wavelengths such as HDF-S and
Chandra.

While from the scientific point of view the EIS strategy has been
quite sensible, from the technical point of view it has been very
demanding as they have involved observations using different
telescope/detector combinations (EMMI, SUSI2, SOFI at the NTT and WFI
at the 2.2m MPG/ESO telescope), a variety of filters, different
surveying strategies, and fields at high- and low-galactic latitudes,
representing different challenges for the astrometric calibration.  To
complicate matters even further EIS observations were conducted in
visitor mode and started right after the commissioning of these
instruments, before their operation had matured.

The short-comings of the first EIS pipeline implementation became
evident when: different telescope/detectors were used (\eg~infrared; CCD
mosaics); the data volume greatly increased due to the use of a $8K
\times 8K$ wide-field imager; and the status of the program changed from
a short-term experimental program to a long-term one.  Using different
telescope/detector combinations and survey strategies entailed the
need to handle: optical and infrared data, requiring different
reduction strategies; detectors with different characteristics;
different types of images and headers; and different combination of
optics, CCDs and filters. The large data volume required the
installation of DLT juke-boxes and hence the need for the dynamical
administration of the machine resources and for procedures to
efficiently store/retrieve data. Finally, the long-term status of the
program combined with the large turnover of visitors (over 25 in three
years, averaging 5-6/year) demonstrated the necessity of simplifying
operations, eliminating as much as possible the need for human
intervention. Clearly, the operational requirements and the timetable
of the program forced the development of a more sophisticated survey
system in order to make the EIS effort sustainable in the long-term.

\section {EIS Survey System Design}

\subsection {Basic Requirements}

Over the past year the EIS team has concentrated its efforts on
designing and implementing a new survey system matched to the
requirements discussed above and building on the hands-on experience
accumulated from the previous surveys . The main goal has been to
achieve a stable and reliable operation by building a close-loop
system providing the tools required for preparing, monitoring and
reporting the progress of the observations, reducing data, preparing
and verifying survey products and releasing them following standard
procedures to ensure efficiency, consistency and quality of the
data. A top requirement has been to make the survey system capable of
handling data from different telescope/detector configurations, as
required by EIS, and of administrating the system resources
automatically.

\void{capable of: preparing the observations based on a well-defined
strategy; monitoring their progress and the overall performance of the
survey; periodically reporting to the community the progress of the
survey; sustaining a large throughput of data with appropriate log and
error reporting; dealing with different sky surveying modes and
telescope/detector configurations; reducing raw images and
subsequently astrometrically and photometrically calibrating them;
preparing image mosaics; producing single-passband source catalogs and
verifying them; producing source catalogs from the combination of data
from different passbands and telescopes; incrementally releasing fully
calibrated images and source catalogs; identifying astronomical
objects of interest; and ensuring the reproducibility of the results
by versioning all software, configuration files and products.}

In addition, the desire to make the system more widely accessible also
required the survey software to: 1) support different users
(\eg~survey/guest); 2) support multi-users and multi-sessions; 3)
allow the user to customize the processing steps and the configuration
files associated to these processes; 4) hide from the user details of
where the data are stored in the system and how they are retrieved; 5)
allow for the installation of the pipeline in other locations; 6)
allow for the easy installation of user-developed algorithms.

In response to this challenge a new fully integrated, end-to-end,
user-friendly pipeline has been designed to make it possible for a
small team to conduct public surveys on a routine basis satisfying the
wide range of requirements outlined above.  While the public surveys
conducted by ESO have so far been relatively modest in size, with VST
and VISTA the trend is for the flow of data to increase
significantly. Therefore, the implementation of a more comprehensive
survey system is essential not only to deal with the present surveys
but also a key element in the long-term plan to ramp-up the system to
cope and fully exploit these new facilities.

\subsection {EIS Graphic Interface}

Taking all the requirements mentioned above into consideration a
graphic user interface was created, consolidating under it all the
functionalities of the pipeline software, making them easily
accessible and their operation simple. To that end all the EIS
pipeline processes have been wrapped by Python scripts which are
launched through this multi-layered graphic interface. Though still
under development a preview of the functionalities of the pipeline is
presented below and illustrated by a development version of the
graphic interface.

\begin{figure*}
\centering
\includegraphics[width=\textwidth]{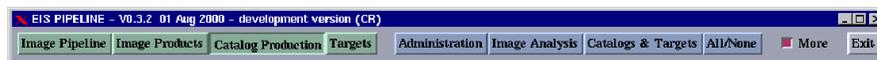}
\caption[]{EIS pipeline graphic interface: top level}
\label{fig:guitoplevel}
\end{figure*}

Basically, the pipeline, represented by different sets of panels,
consists of three layers reflecting the level of complexity of the
processes available.  At the top level there are master-processes that
work as black-boxes, wrapping together all the steps involved in the
production of a given final product. These processes are started by
pressing one of the buttons displayed on the left side of the window
shown in Figure~\ref{fig:guitoplevel}, the first to appear after a new
pipeline session is started. This opens a new window from which these
high-level operations are controlled.  The main processes shown in the
panel of Figure~\ref{fig:guitoplevel} drive: the reduction of raw
images all the way to the production and verification of single
passband catalogs on a run basis; the preparation of final images by
stacking or mosaicing pre-reduced images on a survey/passband basis;
the production of either single passband or color catalogs from
pre-reduced images; and the identification and classification of
objects from the analysis of multi-color catalogs. Under normal
circumstances most survey operations are executed from this layer of
the pipeline.

\begin{figure*}
\centering
\includegraphics[width=\textwidth]{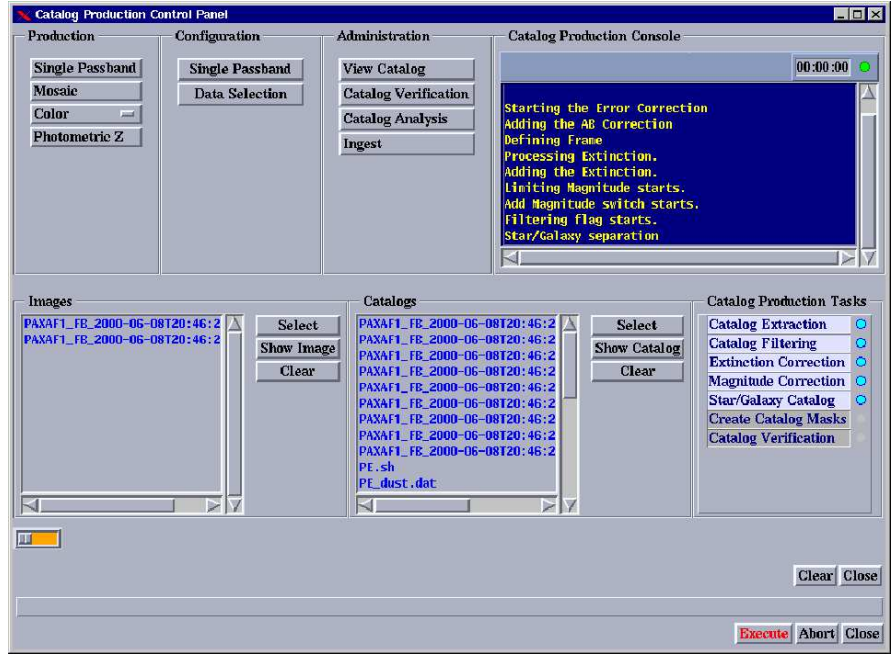}
\caption[]{Catalog production control panel illustrating the steps
involved in the creation of a single passband catalog.}
\label{fig:catprod}
\end{figure*}

As an example, Figure~\ref{fig:catprod} shows the control panel for
catalog production. This panel provides the environment for the
preparation of single or multiple catalogs of different types. It
enables the user to configure the processing steps, the configuration
parameters for each process and to select the appropriate input data
(images or catalogs). The choices include the production of single
passband catalogs, catalogs extracted from an image mosaic, color
catalogs from available single passband catalogs or images, and
photometric redshift catalogs created from color catalogs. This panel
is used after final image products have been produced.

\begin{figure*}
\centering
\includegraphics[width=0.7\textwidth]{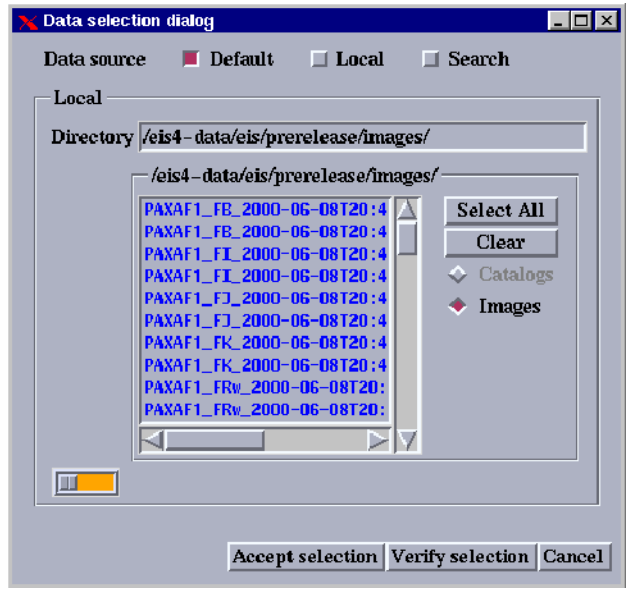}
\caption[]{Graphic interface to access data from different places
including: user defined default directory, local temporary directory
or via the search engine. This interface is common to all panels. The
type of data to be retrieved is automatically set by the type of
product being created.}
\label{fig:dataaccess}
\end{figure*}

\begin{figure*}
\centering
\includegraphics[width=\textwidth]{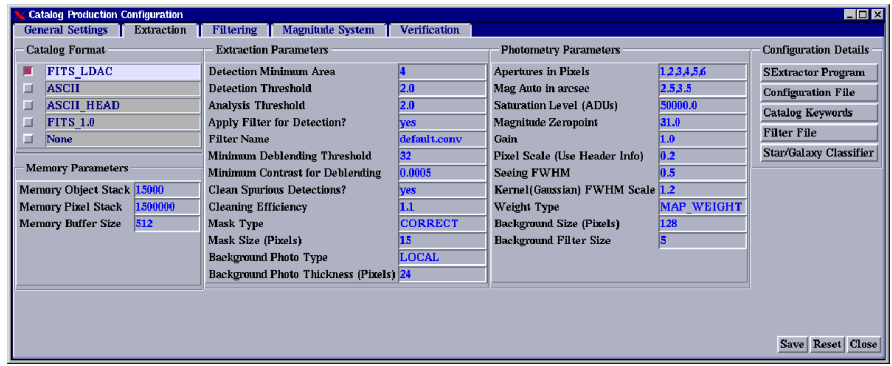}
\caption[]{Example of the  configuration panel associated with the
production of a single passband catalog. From this panel the user can
select the processing steps, change the configuration files associated
with each process and save the results.}
\label{fig:config}
\end{figure*}

Once the end-product type is selected the bottom-half of the panel is
displayed with a layout that depends on the selection made, and the
appropriate links to a so-called data access layer
(Figure~\ref{fig:dataaccess}) and to a configuration panel
(Figure~\ref{fig:config}) are set. The data access layer enables the
retrieval of input files from default, user-specified directories or
via the search engine, which are set for the appropriate type of input
data (catalogs or images). The retrieved data are then displayed in
the list box on the left of the panel, while the one to the right
lists intermediate and final products. The process can be executed in
its entirety or by selecting the steps listed on the right side of the
panel, right below the console (see section~5). The steps to be
executed, the parameters for the different processes and where the
results should be stored can all be set by using the catalog
production configuration window. A console is also available for
monitoring in real-time the progress of the execution. Process logs,
error reports and diagnostic plots are also created and can be viewed
either interactively or at a later time.  While the design of these
graphic interfaces is not final, this example serves to illustrate the
underlying philosophy being adopted throughout the pipeline. It also
underscores some general features of the pipeline among which are the
freedom it gives the user to configure processes and parameters, the
ability of running un-supervised or interactively and to inspect the
results of a process during or after its execution.

\begin{figure*}
\centering
\includegraphics[width=\textwidth]{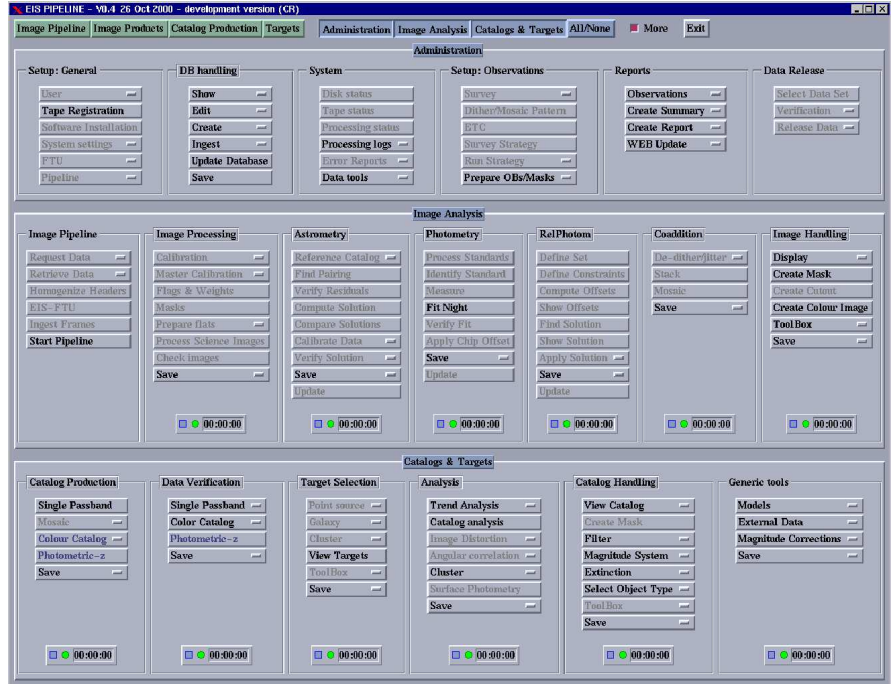}
\caption[]{EIS pipeline GUI: intermediate level}
\label{fig:allpanels}
\end{figure*}

The second layer of the pipeline allows the user to execute step by step
the various processes wrapped together in the first-layer. This layer is
useful for development, testing, training and for setting up suitable
default configuration files. It is also used for the administration of
the pipeline and survey. The main panel is divided into three major
sections: administration, image analysis and catalogs \& targets which
can be accessed individually, using the appropriate buttons on the right
of Figure~\ref{fig:guitoplevel} or all at once as shown in
Figure~\ref{fig:allpanels}. This figure gives a global view of the
functionalities available in the EIS pipeline. Altogether this panel
integrates over 200 processes ranging in functionality and complexity.

The top set of sub-panels are used for the administration of the
pipeline and its database (the three sub-panels on the left) and
surveys (three sub-panels on the right). The administration of the
survey system requires information about users, incoming tapes,
software packages used and their versions, definition of environment
variables, logical devices and paths, and default settings for the
pipeline.  This information is stored in the database via graphic
interfaces to forms that guide the user for the required
information. Also available is a set of graphic interfaces to the
database which is used for development and maintenance. The main
survey database currently consists of over 100 tables. A replica of
these tables are also available for guest users. Finally, disk space,
juke-boxes, processing status of all the EIS machines, logs and error
reports can be monitored and examined directly from the panels.

The three sub-panels on the right are used to administrate long-term
surveys. These sub-panels are used to store in the database
information regarding the instrumental setup (telescope, detectors,
filters) used in a particular survey and their physical
characteristics which are used consistently throughout the
pipeline. They are also used to store information about the regions of
sky to be observed, the pointings and the sequence of exposures
required by the adopted survey strategy. In addition, observing logs
are ingested and night/run summaries and reports are prepared even
before the data are processed. This information is then stored in the
database and used by the system to request data from the archive and
to select runs to be reduced. It is also used to monitor the progress
of the survey and to prepare new observations based on the results of
previous ones. Finally, the procedure to prepare a data release is
also foreseen to be consolidated in this panel. The steps include the
selection of data to be released, final verification of the products,
backup of the particular set of data being released, update of the
database with information about the release, transfer of the data to
the ESO archive and the update of data request page on the web. The
main goal has been to establish standard procedures for all the tasks
of the survey team, hence facilitating their execution.

The middle panel provides access to the different stages of the image
analysis pipeline assembled together in the top layer described above.
The various sub-panels represent different possible breakpoints of the
process, which also correspond to places where data in different
stages can enter the pipeline. This is useful when re-processing of
the data is required. This feature has been extensively used in the
development and test phase. From each sub-panel the specific type of
input data can be retrieved through the data access layer described
earlier. Although this layer of the pipeline is not expected to be
used in routine survey operations, it is very useful for diagnostics,
debugging and bench-marking.

The bottom panel handles the preparation of derived products including
catalog production, target identification and their verification as well
as other analysis algorithms and preparation of simulated data. A
variety of tools are also available to manipulate, visualize and compare
catalogs, correlate objects with information available in other
databases, compute transformations to other photometric systems, compute
extinction corrections, and compare derived results with model
predictions and empirical data.

Finally, at the lowest level, not shown here, tools to handle images
and catalogs in a variety of ways are available individually. These
tools can be found in the sub-panels image and catalog handling as
well as in their respective ``toolboxes''. These toolkits are a
collection of specific software developed over the years to manipulate
pixel maps and catalogs and include the most basic tasks used to build
more complex routines.  Their inclusion into the panels is important
for debugging and testing new versions, before making upgrades.

\subsection {Other Developments and Current Status}

Since a survey system comprises innumerous components the pipeline
software development involved the participation of a large number of
people with different backgrounds and expertises to develop the
variety of algorithms required and to address the vast range of
technical issues involved. Besides the overall framework presented
above other critical areas which required considerable effort have
included: 1) a control system to administrate the flow of the data
through the pipeline and the system resources, to ingest information
into the database and to handle the DLT juke-boxes; 2) a comprehensive
graphic interface to Sybase, the commercial database being used by the
pipeline; 3) the implementation of a search engine, supporting
different combinations of queries; 4) an interactive graphic facility
created by interfacing Python and IDL; 5) a graphic user interface for
the administration of the observed standard stars, the photometric
calibration of the nights and storage of this information; 6) a new
strategy for reducing optical/infrared images, coadding and
astrometrically calibrating them; and 7) algorithms and tools to
verify the data products (see next section).

Other features of the survey system are worth mentioning. First, when
a new session of the pipeline is started, temporary directories
associated to each sub-panel are created and assigned to a
session/user. However, in order to preserve the available disk space,
these directories are automatically removed at the end of the
session. Therefore, the results stored on the directories must be
explicitly saved on permanent directories, tape and database. Second,
all configuration files are identified by a
session/user/process/instrument attribute thus allowing each user to
have its own set of customized configuration files.  Third, all
processes have a common data access layer which can retrieve/store
data from tape/disk and access different databases depending on the
privileges of the user. This ensures the integrity of the survey data
at all times.  Finally, the open nature of the system enables the
system to evolve, making it easy to integrate new tools developed
either by the survey team or by other users.

In summary, the framework provided by the graphic user interface
represents the consolidation of the experience accumulated over the
past few years. the integration of all functions of the survey system
into a single environment and the standardization of the procedures
involved in the execution, reduction, verification and release of data
from a public imaging survey. Currently, the challenge is to integrate
and interface the various pieces of software described above into the
overall pipeline framework. While nearly all the tools have already
been developed and tested their integration into the pipeline is still
in progress.

\section {EIS Products}

Given the nature of the EIS project and the fact that current
technology makes the widespread distribution of pixel maps very
difficult, EIS was charged to prepare source catalogs and to select
astronomical objects of potential interest for VLT
observations~\cite{olsen},\cite{zaggia}. In order to do that in a
well-defined and consistent way, standard procedures and tools have
been developed to prepare and to verify: single passband catalogs;
catalogs extracted from large area mosaics; color catalogs, produced
either by association of single passband catalogs or by using a
reference image~\cite{dacosta},~\cite{benoist}; and
photometric redshift catalogs. These procedures have been extensively
tested and potential problems identified based on the vast experience
gained in processing large amounts of data.

Even in the simplest case of single passband catalogs the procedures
developed extend well beyond the simple extraction of sources and
measurement of flux and structural parameters. The catalog production
includes the standard removal of sources flagged by SExtractor,
trimming of the catalog according to the weight map of its associated
image, extinction correction, conversion of instrumental magnitudes
into a standard system consistent with the filter used in the
observations, automatic masking of regions surrounding bright stars,
and proper consideration of noise properties of the image. The same is
true for color catalog production where proper monitoring of multiple
associations and the accuracy of the relative astrometry are
monitored, depending on the method chosen for its production
(association- or $\chi^2$-method). It is also possible to select the
data for a given passband to serve as the astrometric reference and
redo the astrometric calibration on-the-fly to further ensure a good
relative astrometric solution. Catalogs with photometric redshifts
have also proven to be a valuable tool to identify potential problems
with multi-color data and are produced whenever possible.

\begin{figure*}
\centering
\includegraphics[width=0.75\textwidth]{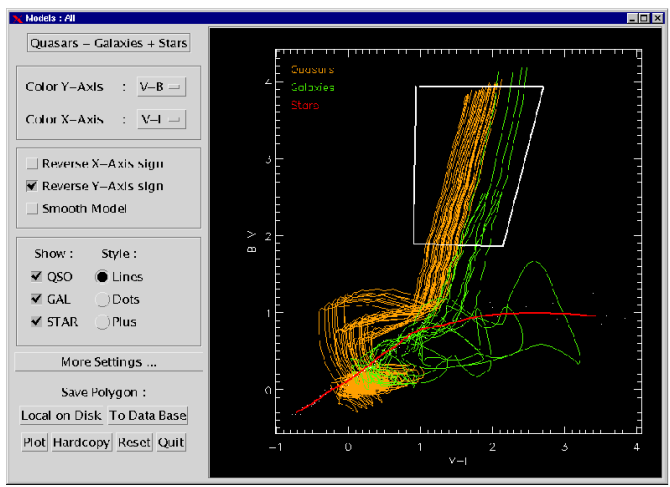}
\caption[]{Color-color diagram showing the expected stellar,
high-redshift galaxies and QSO tracks for a particular combination of
$BVI$ filters.  The $BV$ filters are those used with WFI and the
$I$-filter corresponds to that used in the EMMI@NTT observations}
\label{fig:model}
\end{figure*}

In order to validate the derived catalogs a suite of algorithms and
visualization tools have been developed to examine the catalogs
produced by the pipeline. These tools include: projected distribution,
galaxy and stellar counts, counts-in-cells, angular correlation
function, PSF analysis, color distribution, color-magnitude diagrams,
color-color diagrams, and photometric redshift distribution. Tools are
also available for comparison with other empirical data and/or with
model predictions. To ensure consistency a library of stellar, galaxy
and quasar spectra is available which allows the stellar and
evolutionary tracks to be computed based on information available in
the database regarding the setup used for obtaining the particular
combination of data being verified.

\begin{figure*}
\centering
\includegraphics[width=0.5\textwidth]{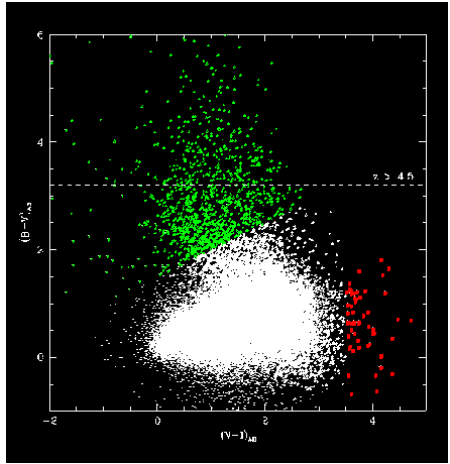}
\caption[]{Example of the output of the application of a "color mask"
created using the tool shown in Figure~\ref{fig:model} applied to the $BVI$
data available over 4 square degrees from the wide-angle surveys.}
\label{fig:ccgal}
\end{figure*}

\begin{figure*}
\centering
\includegraphics[width=\textwidth]{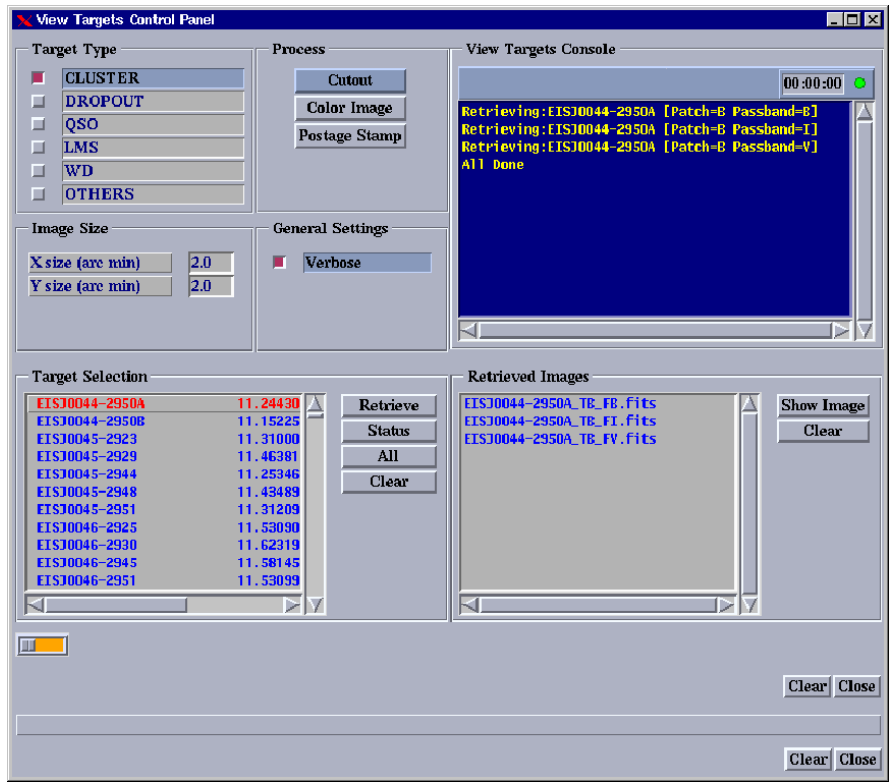}
\caption[]{GUI for inspecting targets selected by the pipeline and
stored in the database. The panel provides different ways of
visualizing the data, to append comments and accept/reject the
object.}
\label{fig:viewtarget}
\end{figure*}

\begin{figure*}
\centering
\includegraphics[width=0.5\textwidth]{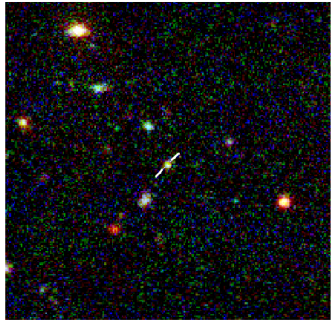}
\caption[]{Example of a color selected target seen in a composite
color image. As explained in the text this tool is useful for
inspecting color catalogs and color-selected targets.}
\label{fig:qso}
\end{figure*}

Tools are also available to search for subsets of objects of potential
interest using selection criteria adopted by other authors or defined
on-line using tools available in the pipeline which allow to define
particular regions of interest, as illustrated in
Figure~\ref{fig:model}. The selected regions can then be stored in the
database and applied to the appropriate color catalogs to locate and
produce a list of low-mass stars, white-dwarfs and high-redshift
objects candidates as well as outliers in the color-color
diagram. Figure~\ref{fig:ccgal} shows the final result of this
analysis for faint sources within a region of 4 square degrees covered
by EIS-WIDE and the Pilot Survey in $BVI$. Besides the expected cloud
of points corresponding to the locus of low-redshift galaxies of
different morphological types, one finds potentially interesting
populations in different parts of the diagram. Note the large number
of high-redshift objects with estimated $z>4.5$.  Comparing the number
of identifications with those predicted by models one may roughly
assess the consistency of the catalog. Furthermore, by examining the
location of outliers in the color image it may be possible to identify
either objects of real scientific interest or possible problems in the
production of the catalog. These selected objects can be stored in the
database and visually inspected using the tool shown
Figure~\ref{fig:viewtarget}, which allows the user to display the data
in different ways such as: cutouts extracted from the images available
in each passband; a composite of postage stamps from all the available
images; a color composite image marking a specific
(Figure~\ref{fig:qso}) or a set of objects of a given type. The panel
shown in Figure~\ref{fig:viewtarget} also allows comments to be
appended and to flag dubious objects. This information can also be
stored in the database and can be used to eliminate objects in the
preparation of final lists of candidates. By examining the objects in
their image context it is possible to detect gross errors in the
preparation of the color catalogs as well as cases where the colors of
the object are likely to be affected by nearby objects.

\section {Summary}

By now the need for public surveys is generally recognized both as a
scientific tool in its own right and as the means to produce a
database of astronomical targets for the increasing number of
large-aperture telescopes already in operation. Perhaps what is not
yet fully appreciated is the magnitude of the effort required to
produce an end-to-end survey system satisfying all the requirements
for a stable and reliable operation of high-throughput running
un-supervised for long stretches of time and making the data and
information available to the community in a timely manner. Building a
proper framework for carrying out such programs is essential to meet
the expectations of the community and win their continuing
support. The survey system presented here provides a solid basis from
which one can ramp-up to handle the significantly larger data rates
expected from future imaging surveys using larger CCD mosaics. On the
short-term it represents a major step towards the ultimate goal of
making the ongoing public surveys at ESO a routine operation,
producing a steady stream of images and derived products that can be
used to prepare statistical samples for VLT programs. Furthermore, in
order to validate these products tools have been developed to handle,
visualize and analyze multi-color data, to compare results with models
or other empirical data and to correlate the derived catalogs with
other databases. These tools are a modest first step towards the
implementation of a more comprehensive archival research environment,
along the lines envisioned by the ongoing effort to build virtual
observatories, as required for the full scientific exploitation of the
data from ongoing and future public surveys.

\bigskip

\noindent{\bf Acknowledgments}\\

I would like to thank all the people involved in the EIS project, in
particular those in the present team that have made significant
contributions to the software development: S. Arnouts, C. Benoist,
C. Rite, M. Schirmer, R. Slijkhuis and B. Vandame. Special thanks to
E. Deul, R. Hook, R. Rengelink and A. Wicenec for their contributions
over the years.

\clearpage
\addcontentsline{toc}{section}{Index}
\flushbottom
\printindex

\end{document}